\documentclass[journal]{IEEEtran}
\ifCLASSINFOpdf
   \usepackage[pdftex]{graphicx}
\else
\fi
\usepackage{amsmath}
\usepackage{cite}

\begin{document}
%
\title{DeepSUM: Deep neural network for Super-resolution of Unregistered Multitemporal images}
%
%
%

\author{Andrea~Bordone~Molini,
        Diego~Valsesia,
        Giulia~Fracastoro,
        and~Enrico~Magli
\thanks{The authors are with Politecnico di Torino -- Department of Electronics and Telecommunications, Italy. email: \{name.surname\}@polito.it. This research has been funded by the Smart-Data@PoliTO center
for Big Data and Machine Learning technologies.}}

%
%

\markboth{}%
{Bordone Molini \MakeLowercase{\textit{et al.}}: DeepSUM: Deep neural network for Super-resolution of Unregistered Multitemporal images}
%



\maketitle

\begin{abstract}

Recently, convolutional neural networks (CNN) have been successfully applied to many remote sensing problems. However, deep learning techniques for multi-image super-resolution from multitemporal unregistered imagery have received little attention so far. 
This work proposes a novel CNN-based technique that exploits both spatial and temporal correlations to combine multiple images. This novel framework integrates the spatial registration task directly inside the CNN, and allows to exploit the representation learning capabilities of the network to enhance registration accuracy. The entire super-resolution process relies on a single CNN with three main stages: shared 2D convolutions to extract high-dimensional features from the input images; a subnetwork proposing registration filters derived from the high-dimensional feature representations; 3D convolutions for slow fusion of the features from multiple images. The whole network can be trained end-to-end to recover a single high resolution image from multiple unregistered low resolution images. 
The method presented in this paper is the winner of the PROBA-V super-resolution challenge issued by the European Space Agency.

\end{abstract}


%
\IEEEpeerreviewmaketitle

\section{Introduction}

Super-resolution (SR) techniques reconstruct a high-resolution (HR) image from one or more low-resolution (LR) images. Remote sensing is playing a key role in mapping and monitoring the Earth and increasing the availability of high spatial resolution data is crucial for many applications such as urban mapping, military surveillance, intelligence gathering, disaster and vegetation growth monitoring. The ever increasing spatial and spectral resolution of instruments onboard of satellites generates large amounts of data which challenge compression algorithms \cite{Valsesia2014TGARS,Valsesia2016Universal} to meet the available downlink bandwidth. This often results in reduced availability of HR products. Combining this issue with the high cost of hardware for smaller missions, it is clear that developing a new generation of post-processing techniques to enhance the spatial resolution is a critical objective.
 
The approaches to image super-resolution can be broadly framed into two main categories: single-image SR (SISR) and multi-image SR (MISR). SISR exploits spatial correlation in a single image to recover the HR version. However, the amount of information available in a single image is quite limited as some information has inevitably been lost in the LR image formation process. Certain applications provide multiple LR versions of the same scene to be combined by means of MISR techniques, where the reconstruction of high spatial-frequency details takes full advantage of the complementary information coming from different observations of the same scene.

For remote sensing problems, multiple images of the same scene can typically be acquired by a spacecraft during multiple orbits, by multiple satellites imaging the same scene at different times, or may be obtained at the same time with different sensors. In this context, developing a successful MISR model hinges on solving important problems such as image registration, invariance to absolute brightness variability, time-varying scene content (e.g., due to the time elapsed between multiple acquisitions), and unreliable data (e.g., due to cloud coverage). Deep learning methods have been proved highly successful in the SISR problem but little work has been done for the MISR problem with remote sensing data.

In this paper we present a deep learning architecture addressing MISR applied to a novel dataset provided by the European Space Agency's Advanced Concepts Team in the context of a challenge \cite{web:kelvins}. The goal of the challenge is to super-resolve images from the PROBA-V satellite. The method presented in this paper won the challenge by achieving the highest fidelity on the reconstructed images.
The unique feature of this dataset is that both LR and HR images have been acquired by the same spacecraft, as opposed to previous works where LR images are artificially down-scaled, degraded and shifted versions of an HR image. The images are not simultaneously acquired so temporal variations exist and have to be handled as well in the super-resolution process.

Our main contribution is DeepSUM, a novel CNN-based architecture to combine multiple unregistered images from the same scene exploiting both spatial and temporal correlations. Our method includes image registration inside the CNN architecture, as a subnetwork named RegNet, which dynamically computes custom filters and applies them to higher dimensional image representations. This is in contrast with the vast majority of deep-learning MISR methods in literature \cite{superdeep} that compensate for the motion as a preprocessing step. This approach allows the registration task to leverage the feature learning capabilities of the network in order to be more accurate and resilient to scene variations, and it also optimizes it in an end-to-end fashion for the final goal of reconstructing a single HR image. The proposed method is blind to the image degradation model as it does not require to explicitly model the blur kernel or the noise statistics, and it is robust to temporal variations in the scene as well as occlusions due to cloud coverage. The only assumption of our model is the translational nature of the shift among LR images.  

A preliminary version of this work \cite{Bordone_Proba-V} addressed MISR for the PROBA-V dataset. 
With respect to our previous work, in this paper we add the registration as part of the network, we improve the use of the image mask information, we expand the experimental results and comparisons with alternative methods and we discuss how a variable number of LR images can be used while the network is designed to handle a fixed input size.

The remainder of this paper is organized as follows. Section \ref{sec:relatedWork} introduces related works on SISR and MISR. Section \ref{sec:challenge} provides details on the novel PROBA-V dataset. Sections \ref{sec:method} and \ref{sec:training} detail the proposed framework and the training procedure. Section \ref{sec:results} contains results and performance evaluation. Section \ref{sec:conclusions} draws some conclusions.

\section{Related work}
\label{sec:relatedWork}
The literature on SR techniques is extensive, both for SISR and for MISR techniques.
SISR approaches can be classified into three main classes: interpolation-based methods (e.g., Lanczos kernels), optimization-based methods and learning-based methods.
Optimization-based methods explicitly model prior knowledge about natural images to regularize this ill-posed inverse problem, and include low total-variation priors \cite{ng2007total},  gradient-profile prior \cite{articleSun2010,6414620} and non-local similarity \cite{4694003,6241428,Zhang2013SingleIS}. Adding prior knowledge restricts the possible solution space generating higher quality solutions. However, the performance of many optimization-based methods degrades rapidly when the upscaling factor increases, and these methods are usually computationally expensive.

Learning-based methods can be pixel-based or example-based. The latter ones are the most popular and they model the correspondence among LR and HR patches for HR patch prediction. After the early work by Freeman et al. \cite{988747} based on searching $k$-nearest neighbors LR-HR patch pairs of the input LR patch to estimate the HR patch, neighbor embedding \cite{1315043,6166881,Bevilacqua2012LowComplexitySS}, sparse-coding \cite{5466111,6175956,5396341,6392274,10.1007/978-3-642-27413-8_47}, anchored neighborhood regression \cite{6751349}, and random forest \cite{7299003} methods were proposed.
More recently deep convolutional neural networks (CNNs) \cite{DnCnnZhang, liu2018non,10.1007/978-3-319-10593-2_13,kim2016deep,kim2015deep_rec,Shi2016RealTimeSI,Lim2017EnhancedDR,Zhang2018ResidualDN} achieved state-of-the-art results for the SISR task. The deep learning paradigm gained attention due to its natural capability of extracting high-level features from images. This is particularly important in remote sensing scenarios where images are highly detailed and their statistics can be very complex.

While most of the deep learning SISR works are related to traditional natural images, lately CNNs have been exploited for remote sensing imagery.
A deep learning based method has been applied by Ma et al. \cite{8600724} on remote sensing images in the frequency domain. Their CNN takes as input discrete wavelet transformed images and adopts recursive block and residual learning in global and local manners to reconstruct HR wavelet coefficients. 
Jiang et al. \cite{8677274} proposed a generative adversarial network-based edge-enhancement network for robust satellite image SR reconstruction.

Concerning MISR, the first work was proposed by Tsai and Huang \cite{tsaiHuang1984}, who used a frequency-domain technique to combine multiple under-sampled images with sub-pixel displacements to improve the spatial resolution of Landsat TM acquisitions. 
Due to the drawbacks of the frequency-domain algorithms, like the difficulty to incorporate the prior information about HR images, many spatial-domain MISR techniques were proposed over the years \cite{935034}.
Typical spatial-domain methods include non-uniform interpolation \cite{1176931}, iterative back-projection (IBP) \cite{IRANI1991231}, projection onto convex sets (POCS) \cite{Stark:89,413332}, regularized methods \cite{1331445,4060955,shen2009}, and sparse coding \cite{Kato:2017:DSM:3066426.3066466, KATO201564}.

The iterative back projection (IBP) approach was introduced by Irani and Peleg \cite{IRANI1991231}. IBP aims to improve an initial guess of the super-resolved image by back projecting the difference between simulated LR images and actual LR images to the SR image. The updates are iteratively performed attempting to invert the forward imaging process. Drawbacks come from the inability to deal with  unknown or very difficult to model image degradation processes, as well as the difficulty in including image priors. 

Another class of MISR methods is the projection onto convex sets (POCS) \cite{Stark:89,413332}, where restoration and interpolation problems are simultaneously solved to estimate the SR image, after accurate motion compensation. 
Despite allowing an easy incorporation of a priori knowledge, POCS suffers from slow convergence and high computational cost.


Regularized methods are some of the most effective multi-frame SR reconstruction approaches.
In the past decades, many kinds of regularizers have been proposed to preserve edge information while removing image noise, such as Tikhonov regularizer \cite{Hardie98,913592}, Markov random field regularizer \cite{6096366}, total variation (TV) \cite{661187,Marquina2008,Zhang2014} and  bilateral total variation (BTV) \cite{1331445}.
In particular, a few works have been proposed for remote sensing applications. Shen et al. \cite{shen2009} proposed a maximum-a-posteriori (MAP) SR method with Huber prior for MODIS images captured in different dates. 
Another multi-temporal SR method was proposed by Li et al.\cite{5308275} for  Landsat-7 PAN images. 
Instead, other works \cite{6134690,Zhang2014} proposed SR methods for multi-angle remote sensing captures. 

In recent years, sparse coding methods based on dictionary learning have successfully been applied to MISR \cite{Kato:2017:DSM:3066426.3066466, KATO201564}. 

Most of the above SR methods assume a priori knowledge of the motion model, blur kernel and noise level, where both blur identification and image registration are performed as a preprocessing stage before reconstruction. However, there are many applications where knowing the image degradation process or reliably estimating it can be challenging. For this reason, many studies have been carried out on blind SR image reconstruction \cite{941854,He2005ARF}. These blind methods usually work in two stages, namely, (1) image registration from LR images, followed by (2) simultaneous estimation of both the HR image and blurring function. 
In order to reduce the effect of registration errors, some researchers have developed methods to simultaneously estimate the motion parameters and the reconstruction \cite{650116,Zhang2015}. Hardie et al. \cite{650116} presented a  MAP framework to jointly estimate image registration parameters and the HR image. 
Along the same lines, Zhang et al. \cite{Zhang2015} also integrated the joint estimation of the blurring function.
Moreover, Kato et al. \cite{Kato:2017:DSM:3066426.3066466} recently proposed a sparse coding method where image registration and sparse coding are performed in a unified framework reducing the image registration error. 
 
In the last years, deep learning based methods have been proposed to solve similar MISR problems in context of video super-resolution \cite{7444187,DBLP:journals/corr/CaballeroLAATWS16}. Most of these works are composed of two steps: a motion estimation and compensation procedure followed by an upsampling process, heavily relying on the prior motion estimation. Recently, Jo et al. \cite{Jo_2018_CVPR} presented a novel end-to-end residual  CNN  to produce a SR image without explicit motion compensation. A CNN is trained to simultaneously solve motion estimation and HR image reconstruction tasks by producing a set of pixel-dependent filters and a residual correction. A similar idea was developed by Tian et al. \cite{Tian2018TDANTD}. However, little work has been done on deep learning MISR methods in the context of remote sensing, which poses specific challenges. Kawulok et al. \cite{SuperDeep_MISR} propose a MISR method that does not fully exploit the benefit of deep learning, restraining their CNN to solve a SISR problem. The fusion of the upsampled LR images is performed by the median shift-and-add method, generating a SR image that is used as initial guess for a classic regularized method. Their method is not end-to-end trainable in a supervised manner and their CNN is trained against LR images obtained by artificially degrading HR images.
Inspired by the recent video super-resolution works, we aim to tackle the MISR problem on satellite images by jointly registering the input LR images and reconstructing the SR image, all within an end-to-end trainable CNN, where the two tasks are optimized jointly.

\section{The PROBA-V SR dataset}
\label{sec:challenge}

At present, it is difficult to find a dataset collecting both a set of real-world LR observations and the corresponding HR image for the same scene, as captured from the same platform. Many of the works found in the SR literature are based on simulated data, where LR observations for a specific scene are obtained through a degradation and down-sampling process of the HR images by assuming a sensor imaging model. This is a simplified scenario as it either assumes a non-blind problem, i.e., the degradation model can be characterized to some extent, or has the limitation that a too simple degradation model may not accurately match the real one, especially when in presence of temporal variations in the scene content.

The Advanced Concepts Team of the European Space Agency has issued a competition \cite{2019arXiv190701821M} to perform MISR for the images acquired by the PROBA-V satellite. PROBA-V is an Earth observation satellite designed to map land cover and vegetation growth across the entire globe. It was launched in 2013 into a Sun-synchronous orbit at an altitude of 820km. Its payload provides an almost global coverage with 300m LR images and 100m HR images. However, the HR images are acquired with a higher revisit time, roughly one every 5 days, instead of one per day. The dataset gathers satellite data from 74 regions located around the world from the PROBA-V mission. Images are provided as level 2A products composed of radiometrically and geometrically corrected Top-of-Atmosphere reflectance in Plate Carrée projection for the RED and NIR spectral bands. The size of the collected images is $128 \times 128$ and $384 \times 384$ for the LR and HR data respectively. The images have a single channel with a bit-depth of 14 bits. Each data point consists of one HR image and several LR images (ranging from a minimum of 9 to a maximum of 30) from the same scene. In total, the dataset contains 1160 scenes, 566 are from NIR spectral band and 594 are from RED band. 
The images of a specific scene are captured at multiple times over a maximum period of 30 days. Weather and changes in the landscape pose a limitation in the similarity of the images. Clouds, cloud shadows, ice, water, missing regions, presence of agricultural activities and, in general, human activity are the main sources of inconsistency across these images, thus posing a major challenge for any image fusion method.
Moreover, each image comes with a mask, indicating which pixels in the image can be reliably used for reconstruction (e.g., they are not covered by clouds).
The geometric disparity among the images can be considered as translational only. Subpixel shifts in the content of the LR images do occur and are indeed important for the MISR task.

The unique nature of this dataset (with real LR and HR images captured by the same platform at multiple times) makes for an interesting case study for SR techniques. Developing SR products from multiple, more frequent LR images could simultaneously provide enhanced resolution and higher temporal availability and is therefore an interesting application of MISR. Moreover, having real images of the same scene for both the low and high resolutions enables data-driven methods such as CNNs to learn the inversion of possibly complex degradation models and the best feature fusion strategy to handle temporal variations.

\begin{figure*}[t]
\centering
\includegraphics[width=7.1in]{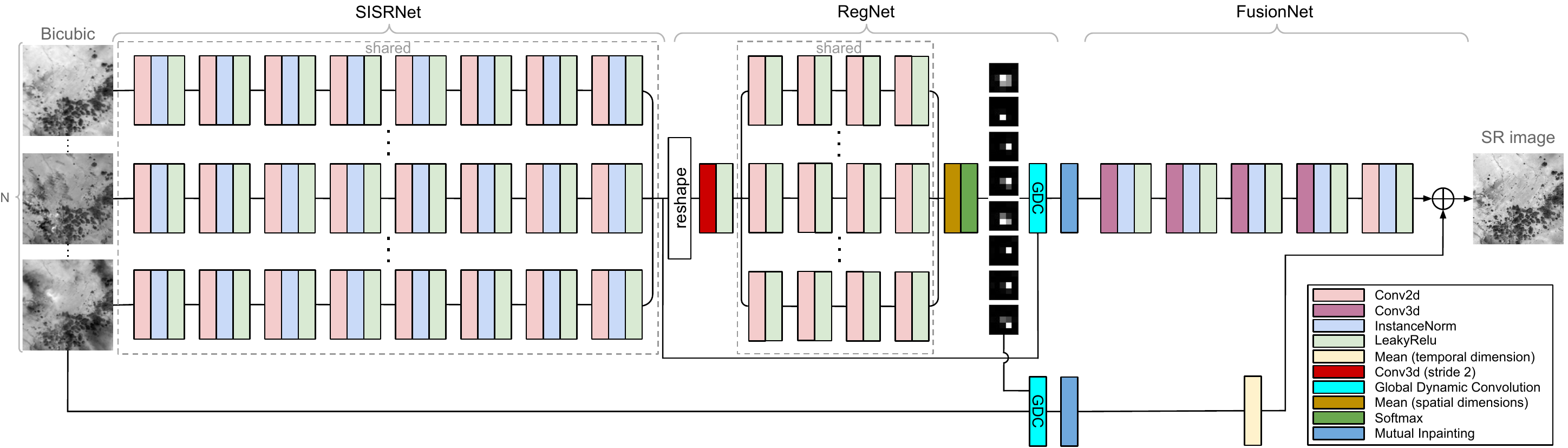}
\vspace{-0.15cm}
\caption{DeepSUM network. The $N$ input bicubic-upsampled and registered images are independently processed by a SISRNet subnetwork, and their features used by the RegNet to compute registration filters to register the feature maps of the $N$ images to each other. The FusionNet subnetwork merges the features of the images to produce a residual image. The residual image is then added element-wise to the average of the registered input to obtain the SR image.}
\vspace{-0.15cm}
\label{fig:Architecture}
\end{figure*}

\section{Proposed method}
\label{sec:method}
The proposed method, called DeepSUM, reconstructs a high-resolution image $I^\text{HR}$ given a set of $N$ LR images  $I_{[0,N-1]}^\text{LR}$ representing the same scene:
$${I}^\text{SR}=f(I_{[0,N-1]}^\text{LR},\theta),$$
where $\theta$ represents the model parameters and $f$ represents the mapping function from LR to HR.
$I_{[0,N-1]}^\text{LR}$ and $I^\text{HR}$ are represented as real-valued tensors with shape $N \times H \times W \times C$ and $1 \times rH \times rW\, \times C$ respectively, where $H$ and $W$ are the height and the width of the input LR frames, 
$C$ is the number of channels and $r$ is the scale factor. 
While the LR images roughly represent the same scene as the HR image, there are several factors to be considered:
\begin{itemize}
    \item the LR images are not registered with each other; 
    \item the LR images and the HR image are not registered;
    \item the brightness of the HR image may be different from that of any LR image;
    \item the scene changes over multiple acquisitions;
    \item LR and HR images may be covered by different clouds and cloud shadows patterns or affected by corrupted pixels. 
\end{itemize}

To tackle this problem we propose to employ a supervised deep learning approach, where a CNN learns the residual between bicubic interpolation and the ground truth.
As a preprocessing step, the LR images are bicubically interpolated to the desired size and then fed into a CNN composed of three main building blocks. An overview of the network is shown in Fig. \ref{fig:Architecture}.

The first block, called SISRNet, is a feature extractor that can be seen as a SISR network without the output projection to a single channel. Each of the $N$ input images is processed independently by a sequence of 2D convolutional layers. The convolutional filters are shared along the temporal dimension, i.e., all the $N$ interpolated LR (ILR) images go through the same set of filters.

The second network block, called RegNet, aims at estimating a set of filters to register the $N$ higher dimensional image representations produced by the SISRNet block to each other at integer-pixel precision (notice that the network is working at the same spatial resolution as the HR image, so integer shifts correspond to sub-pixel shifts in the LR data). RegNet has been devised to align $N-1$ instances with respect to the first, taken as reference, by operating purely translational shifts. Therefore, the output is a set of $N-1$ 2D filters to be applied spatially to each feature map of the $N-1$ inputs. 

Finally, the third block, called FusionNet, merges the registered image representations in the feature space in a ``slow'' fashion, i.e., by exploiting a sequence of 3D convolutional operations with small kernels. The output is a single super-resolved image.

In the following, we are going to describe each individual block more in detail.

\subsection{SISRNet Architecture}

The goal of SISRNet is to exploit spatial correlations to improve upon the initial bicubic interpolation. In doing so, the network learns to extract visual features that can be conveniently exploited by the subsequent network blocks. SISRNet has multiple 2D convolutional layers whose weights are shared among the $N$ input images, effectively processing each of them independently. Each convolutional layer is followed by Instance Normalization \cite{vedaldi2016instance}. Instance normalization is used in place of Batch normalization \cite{ioffe2015batch} to make the network training as independent as possible of the contrast and brightness differences among the input images.

\subsection{RegNet Architecture} \label{sec:regnet_arch}

RegNet is composed of two sub-blocks: a CNN, and a global dynamic convolutional layer (GDC).
The CNN processes the higher dimensional image representations $Z_{[0,N-1]}^\text{ILR}$ generated by SISRNet block and outputs a set of $N-1$ filters $G_{[1,N-1]}$. Each filter $G_i$ is subsequently applied in the spatial dimensions to each of the channels of $Z_i^\text{ILR}$ by the GDC layer by means of a 2D convolution in order to register each feature map of $Z_{i}^\text{ILR}$ with respect to the reference one $Z_{0}^\text{ILR}$. The filters $G_{[1,N-1]}$ have a fixed support equal to $K \times K$ that upper bounds the maximum possible translational shift correction to $\lfloor K/2 \rfloor$. Notice that there is an implicit assumption that all feature maps of an image require the same shift to be registered with the reference, so that the computed filter is shared channel-wise. The registered feature maps $Z_{[0,N-1]}^\text{IRLR}$ of the $N$ images are thus obtained as:

\begin{align*}
G_{i}&=f_\text{RegNet}(Z_{[0,N-1]}^\text{ILR},\theta_\text{RegNet}), \quad i=1,\dots,N-1\\
Z_{i}^\text{IRLR}&= \begin{cases}
Z_{i}^\text{ILR}, \quad i=0\\
G_{i} * Z_{i}^\text{ILR}, \quad i=1,\dots,N-1\\
\end{cases},
\end{align*}
being $*$ the 2D convolution operator. The same filters are also applied to the input ILR images to register them in the residual connection:
\begin{align*}
I_{i}^\text{IRLR}&= \begin{cases}
I_{i}^\text{ILR}, \quad i=0\\
G_{i} * I_{i}^\text{ILR}, \quad i=1,\dots,N-1\\
\end{cases},
\end{align*}

The novelty of this network is twofold: firstly the filters are dynamically computed for each input image, and secondly it makes use of the features to compute the per-image optimal registration instead of performing it in image space, like most of motion estimation algorithms do. This allows to leverage the powerful feature space of the network to boost the registration performance by making it robust to scene variations. In addition, it is fully differentiable so that the whole architecture can be trained end-to-end.

\begin{figure}[t]
\centering
\includegraphics[width=3.2in]{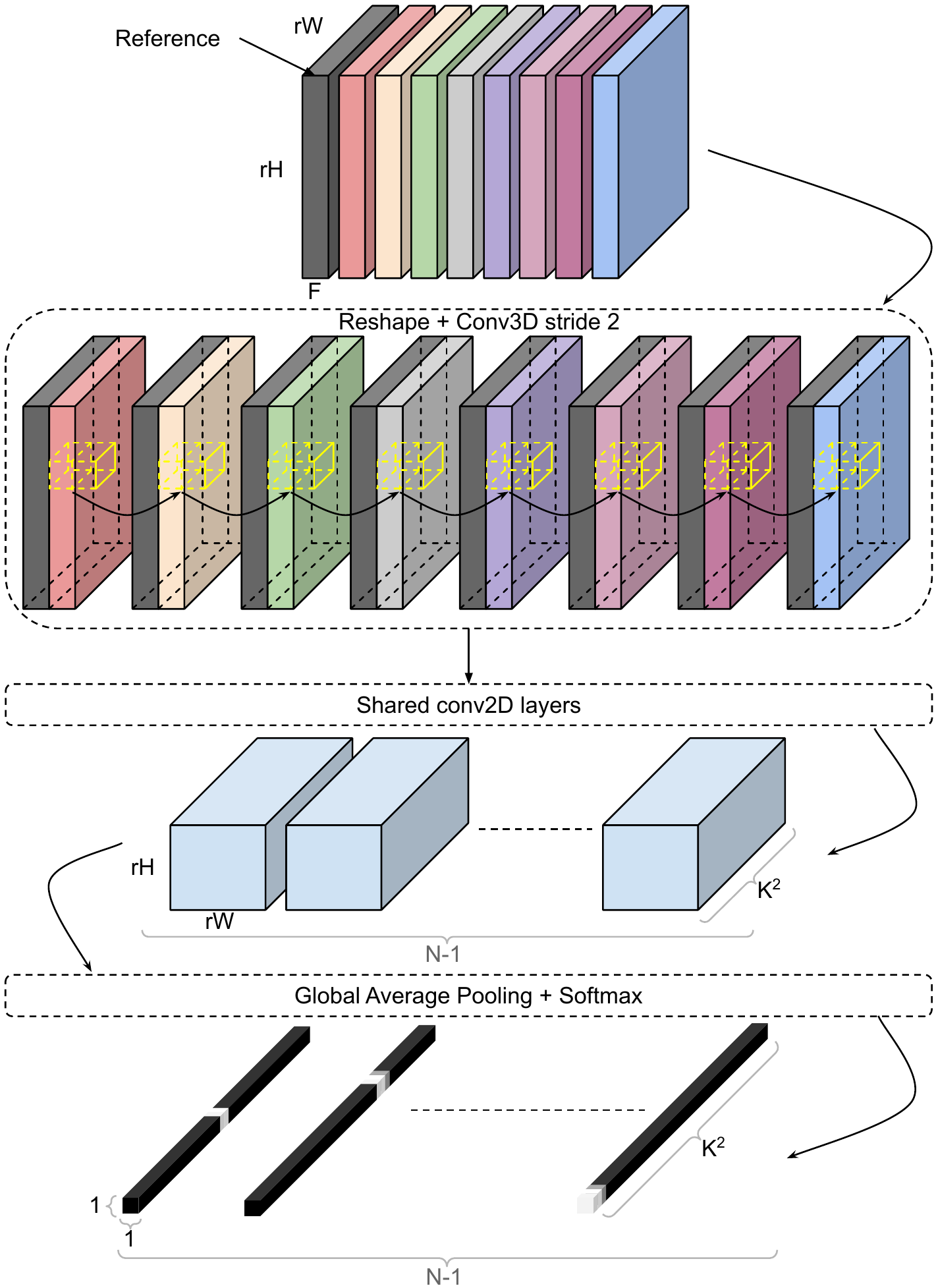}
\vspace{-0.15cm}
\caption{Visual depiction of the RegNet operations to generate the dynamic registration filters from the image features produced by SISRnet.}
\vspace{-0.15cm}
\label{fig:Detailed_Regnet}
\end{figure}

More in detail, the operations performed by RegNet are depicted in Fig. \ref{fig:Detailed_Regnet}. SISRNet outputs a tensor $Z^\text{ILR}$ with shape  $N  \times rH \times rW\, \times F$, where $F$ is the number of features, that is reshaped before being fed to RegNet. The features of the first image $Z_{0}^\text{ILR}$ are chosen as a reference and a new tensor of size $2(N-1)  \times rH \times rW\, \times F$ is built by replicating the reference $Z_{0}^\text{ILR}$ $N-1$ times and interleaving each replica with the other $(N-1)$ image representations $Z_{[1,N-1]}^\text{ILR}$. This sequence of paired reference/unregistered features is then processed by convolutional layers to produce the filters. 
RegNet has a first 3D convolutional layer and a series of shared 2D convolutional layers. The first layer is the key component of registration and it processes the $2(N-1)$ image representations in pairs by using a stride equal to 2 along the temporal dimension and filters of shape $2 \times 3 \times 3$. This operation allows to correlate the features of each  $Z_{i}^\text{ILR}$ with respect to the ones of the reference $Z_{0}^\text{ILR}$ and compute the shift. Notice that this processing in pairs is necessary to avoid any ordering ambiguity and let the network understand that the output is relative to the reference.
After this 3D convolutional layer the output tensor has shape $(N-1)  \times rH \times rW\, \times F$.

This tensor passes through a series of 2D convolutional layers with shared weights along the temporal dimension. 
The last RegNet 2D convolutional layer applies a number of kernels corresponding to the spatial size of the dynamic filters $K \times K$, obtaining a tensor with shape $(N-1)  \times rH \times rW\, \times K^2$. Each value over the spatial dimensions can be seen as a local estimate of the desired shift based on the local image representation. Since there is a global translational shift by assumption, the values are averaged over the spatial dimensions to obtain a tensor with shape $(N-1) \times 1 \times 1 \times K^2$.

\begin{figure}[t]
\centering
\includegraphics[width=3.0in]{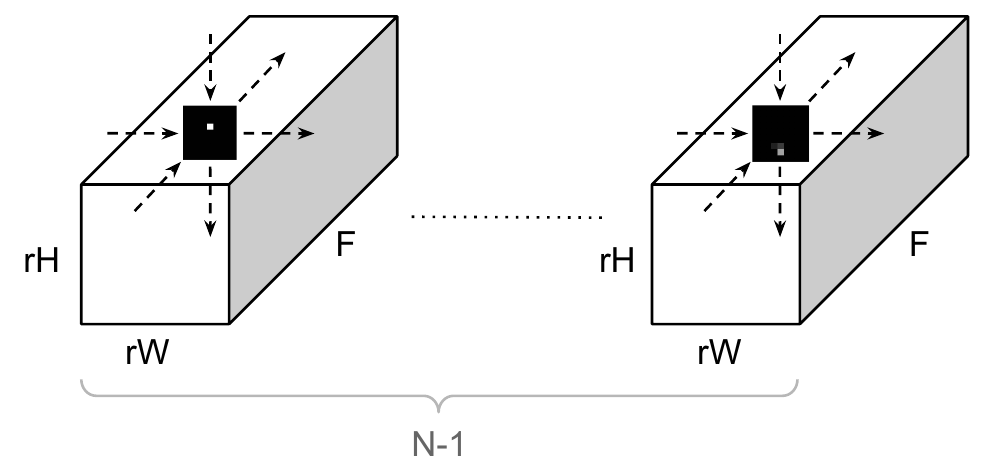}
\vspace{-0.15cm}
\caption{GDC: convolution between the dynamic filters and the image representations to align them with respect to the reference.}
\vspace{-0.15cm}
\label{fig:DC}
\end{figure}

Finally, this tensor is passed through a softmax layer, so that the values over the last dimension ($K^2$) add up to 1. The softmax layer promotes a spiked filter with most elements set to zero \cite{Brabandere2016DynamicFN}.
The final tensor represents the $(N-1)$ dynamic filters with shape $K \times K$ to be used to register the $(N-1)$ image representations with the GDC operation, as in Fig. \ref{fig:DC}.

\subsection{Mutual Inpainting}
The registered and interpolated feature maps $Z_{[0,N-1]}^\text{IRLR}$ have regions with unreliable values due to cloud coverage, shadows, corrupted pixels and so on. 
A per-pixel boolean mask is assumed to be available as side information, with the purpose of mapping pixels that can be reliably used for the fusion task. An example on how to obtain such mask is to run a cloud detection algorithm on the image to segment areas with clouds. This is very important because areas occluded by clouds do not provide any useful information.
In order to prevent FusionNet from combining feature maps from multiple images where some have unreliable intensities, we fill the masked areas with values from the feature maps of other images.
The regions with missing or unreliable values in each feature map of each image are filled with values taken from the corresponding feature map of other images having reliable values in those regions, if any are available. In the case none of the images has feature maps with reliable values, we keep those unreliable regions as they are.
Since after RegNet the masks are not aligned with the corresponding image representations, we shift the masks by an integral shift as close as possible to subpixel shift computed and operated by RegNet.
This procedure is performed on both the residual image representations $Z_{[0,N-1]}^\text{IRLR}$ and the registered input images $I_{[0,N-1]}^\text{IRLR}$ right before averaging them.

\subsection{FusionNet Architecture}

The $N$ registered outputs $Z_{[0,N-1]}^\text{IRLR}$ are progressively fused by the FusionNet subnetwork. FusionNet is composed of $\lfloor N/2 \rfloor$ 3D convolutional layers where convolution is performed without any padding in the temporal dimension, so that the temporal depth eventually reduces to 1. This architecture implements a ``slow'' fusion process in the feature space, which allows the network to learn the best space to decouple image features that are relevant to the fusion from irrelevant variations and to construct the best function to exploit spatio-temporal correlations \cite{DBLP:journals/corr/CaballeroLAATWS16}. 
Finally, the proposed architecture employs a global input-output residual connection. The network estimates only the high frequency details necessary to correct the bicubically-upsampled input. This is an established technique for image restoration problems using deep learning \cite{DnCnnZhang}, including SISR. However, with respect to SISR, our proposed network is a many to one mapping, so the residual is actually added element-wise to a basic merge of $I_{[0,N-1]}^\text{IRLR}$ in the form of their average. Notice that registration of the input images is performed before averaging by means of the same filters produced by RegNet. Hence, the output is computed as follows:
\begin{align*}
   \bar{I}^\text{IRLR} &= \frac{1}{N}\sum_{i\in[0,N-1]}{I_{i}^\text{IRLR}},\\
   I^\text{SR} &= \bar{I}^\text{IRLR} + R.
\end{align*}
being $R$ the residual estimated by the CNN.

\subsection{Loss Function}
\label{subsec:loss}
Model parameters are optimized by minimizing a loss function computed as a modified version of the Euclidean distance between the SR image and the HR target. Minimizing the Euclidean distance is optimal in terms of the mean-squared error metric. Some deep learning works on SISR attempted to use an adversarial loss \cite{ledig2017photo}. While this approach produces visually pleasing results, it tends to hallucinate information, resulting in lower MSE scores and less reliable products in the context of remote sensing; hence, the adversarial approach has not been followed in the present work.
As we mentioned in Sec. \ref{sec:challenge}, since the PROBA-V satellite does not capture LR images and HR images of a specific ground scene simultaneously, there are discrepancies coming from different weather conditions, changes in the landscape and variable absolute brightness due to the large interval between scene acquisitions. The LR images could be quite different from one another and from the corresponding HR image as well. For this reason, we must make the training objective as invariant as possible to such conditions. In particular, in order to build invariance to absolute brightness differences between ${I}^\text{SR}$ and ${I}^\text{HR}$, the modified loss function equalizes the intensities of the SR and HR images so that the average pixel brightness is the same on both images.
Moreover, since ${I}^\text{SR}$ and ${I}^\text{HR}$ could be shifted, the loss embeds a shift correction. ${I}^\text{SR}$ is cropped at the center by $d$ pixels, i.e., as many pixels as the maximum expected shift. Then all possible patches $I_{u,v}^\text{HR}$ of size $(rH-d) \times (rW-d)$ for vertical and horizontal shifts $u,v$ are extracted from the target ${I}^\text{HR}$. All possible Euclidean distances are computed and the minimum one is taken as loss to optimize.
In summary, our loss is as follows:
\begin{align*}
L=\min_{u,v\in[0,2d]}\Vert{I_{u,v}^\text{HR}-({I}_\text{crop}^\text{SR}+b)}\Vert^2,
\end{align*}
where ${I}_\text{crop}^\text{SR}$ is the cropped version of $I^\text{SR}$ and $b$ represents the brightness correction:

$$b=\frac{1}{(rW-d)(rH-d)}\sum_{x,y} \left( I_{u,v}^\text{HR}-{I}_\text{crop}^\text{SR} \right).$$

The loss is computed by utilizing only the HR image pixels that are marked as reliable by the mask provided with the dataset and the SR image pixels for which at least one out of $N$ LR images were clear. The reason for this is that a cloud in the HR image can never be predicted from terrain data in the IRLR images, so its pixels should not contribute to the loss function. Viceversa, it is also impossible to predict HR terrain if all the IRLR images have concealed regions.

\section{Training process}
\label{sec:training}
\subsection{Pre-training}
Training the whole network end-to-end from scratch is hard due to several local minima that do not make SISRNet, RegNet and FusionNet work as expected. For example, the gradients computed during training do not sharply discriminate the RegNet task to generate registration filters from the high-resolution feature learning of SISRNet. 

In order to solve this issue, it is possible to pretrain each block to handle its specific subtask, and then combine all the blocks to be fine-tuned in an end-to-end fashion.

\subsubsection{SISRNet pre-training}

As mentioned in Sec. \ref{sec:method}, SISRNet aims to independently super-resolve each of the $N$ input images, while providing useful higher dimensional image representations. SISRNet is pretrained by setting up a pure SISR problem (i.e., a single input image) where an additional projection layer is added at the end, in order to turn the high-dimensional feature space into a single-channel image. SISRNet with the final projection layer is trained with the same objective function of the final training, where the single image reconstruction is compared with the only HR image available for the scene.
The rationale behind this is to make SISRNet exploit spatial correlations as much as possible to generate the best image features for the SISR task. 
Once the pretraining procedure is completed, the final layer is removed and a dataset of feature maps of the input training images is generated to pretrain RegNet. 

\subsubsection{RegNet pre-training}

The purpose of pre-training RegNet is learning to generate registration filters, i.e., filters that shift the feature maps of the $N-1$ input images with respect to the reference input. This operation would be quite challenging to learn if the whole network was trained end-to-end, so its pretraining is crucial for the overall network performance.
RegNet is pre-trained by casting registration as a multi-class classification problem. Each dynamic registration filter generated by the network is viewed as a probability distribution over the possible shifts with the objective of estimating the correct shift. The number of classes is $K^2$ since the filter size is $K \times K$. In case of an ideal shift of an integer number of pixels, the predicted filter should be a delta function centered at the desired shift.

The input data to be used for the pretraining of RegNet are the feature maps produced by the pretrained SISRNet for the images in the training set. As described in Sec. \ref{sec:regnet_arch}, the input to RegNet are $N$ feature maps from images of the same scene. These feature maps are then synthetically shifted with respect to the first one by a random integer amount of pixels. The purpose is to create a balanced dataset where all possible $K^2$ classes (shifts) are seen by the network. The desired output is a filter with all zeros except for a one in the position corresponding to the chosen shift. A cross-entropy loss between the softmax output and the true filter is used to learn the RegNet weights.

\subsection{Final training}
The proposed network is finally trained as a whole, end-to-end for the MISR task. FusionNet is trained from scratch while SISRNet and RegNet weights are initialized from the pretraining procedures.
The concurrent optimization of all the network blocks allows SISRNet to finetune the image representations to facilitate the RegNet task that in turn finds the best registration to boost the efficiency of FusionNet. 

\subsection{Testing phase} \label{sec:testing}
The network architecture presented in the previous sections has been designed to deal with a fixed number $N$ of LR images for a given scene. However, it might happen that more than $N$ images are available and exploiting them could further boost the SR reconstruction performance. Therefore, during testing, one can perform multiple forward passes by using multiple subsets of the available images. Each subset will produce a different SR estimate and, in the end, all SR estimates are averaged. Notice that the estimates should be registered to each other so it is advisable to always use the same LR image as the reference in the network (e.g., one could choose the image with fewer masked pixels). One method to produce useful subsets when more than $N$ LR images are available is to sort them by increasing number of masked pixels and then use a sliding window over $N$ images to compute SR estimates. It must be remarked that the SR estimate quality degrades with increasing number of masked pixels. Also, the estimates are clearly not independent if some images are reused multiple times, but we found consistent gains on our test set, nevertheless.

Defining the optimal function to merge SR estimates or making the network independent of the number of input images could be studied in future research.

\section{Experimental results and discussions}
\label{sec:results}
In this section we perform an experimental evaluation of DeepSUM, comparing it with several alternative approaches. Code and pretrained models are available online\footnote{https://github.com/diegovalsesia/deepsum}.
We first perform an ablation study to highlight the contribution given by RegNet to the overall network performance.
Then, we assess the performance of alternative approaches.

\subsection{Experimental setting}
In the following experiments, we employ both the NIR and RED band datasets described in Sec. \ref{sec:challenge}. We use 396 scenes for training and 170 for testing from the NIR band dataset and 415 for training and 176 for testing from the RED band dataset. Expanding the training set with more scenes should further improve performance as more variability can be captured by our model. Since DeepSUM is devised to work with a fixed size temporal dimension, we train the network using the minimum number of images available for each scene, i.e., $N=9$ images. When more images are available we select the 9 clearest images according to the masks.  As a preprocessing step, all LR images are clipped to $2^{14}-1$ since corrupted pixels with large values occur in the LR images throughout the PROBA-V dataset. 

After the bicubic interpolation, each scene is a data-cube of size $9 \times 384 \times 384$, from which we extract a dataset with patches of size $9 \times 96 \times 96$. 100 random patches are extracted from each scene, resulting in a total of 38400 samples. The patches are extracted considering the available pixel masks: a patch is accepted only if at least 9 scene images are at least 70\% clear and the HR image in the same coordinates is at least 85\% clear. The amount of unreliable pixels is relaxed to keep as much information as possible from the original images at the cost of training with sub-optimal patches. Separate networks are trained for RED and NIR. The proposed network is trained for around 3000 epochs with a batch size of 8 for both RED and NIR.

The Adam optimization algorithm \cite{kingma2014adam} is employed for training, with momentum parameters $\beta_{1} = 0.9$, $\beta_{2} = 0.999$, and $\epsilon = 10^{-8}$. The learning rate $\lambda$ is initialized to $5\times 10^{-6}$ for the whole network. 
We employ the Tensorflow framework to train the proposed network on a PC with 64-GB RAM, an Intel Xeon E5-2609 v3 CPU, and an Nvidia 1080Ti GPU.
The exact number of network layers is shown in Fig. \ref{fig:Architecture} and the number of filters is 64 everywhere except for the RegNet's first layer, which has 128 filters. In order to mitigate border effects, we use reflection padding in all 2D convolutions.
Each layer in the network is followed by Leaky ReLU non-linearity, except for the last layer. Each layer in SISRnet and FusionNet is followed by an Instance Norm layer.
Instance normalization \cite{vedaldi2016instance} is used in place of Batch normalization layer to make the network training as independent as possible of the contrast and brightness differences among the input images.
Finally, since the network produces a residual estimate $R$, we normalize $\bar{I}^\text{IRLR}$ and $I^\text{HR}$ so that their difference gives a unit variance residual $R$, thus avoiding any scaling to be performed by the last layer of the network and improving convergence speed.

\subsection{Quantitative results}

The evaluation metric that we consider is a modified version of the PSNR (mPSNR), from which we derived the loss function described in Sec. \ref{subsec:loss}.
$$\text{mPSNR}=\max_{u,v\in[0,6]}20\log\frac{2^{16}-1}{\parallel{I_{u,v}^\text{HR}-({I}_\text{crop}^\text{SR}+b)}\parallel^2}$$
The mPSNR computation is meant only for pixels that are not concealed both in the target HR image and in the reconstructed image. Similarly to the loss function during training, this metric has been devised to cope with the high sensitivity of the PSNR to biases in brightness and with the relative translation that the reconstructed image might have with respect to the target HR image. In this case the maximum mPSNR over all possible shifts is considered for evaluation. Note that, by design of the dataset, the maximum shift in the horizontal and vertical directions is equal to 6 pixels.

We remark that this metric was also used to evaluate submissions to the ESA challenge, where the score was computed as a ratio between the mPSNR of the submission and that of the baseline approach, average over all the held-out test set.

\subsubsection{Ablation study}

\begin{figure}[t]
    \centering
    \includegraphics[width=0.43\textwidth]{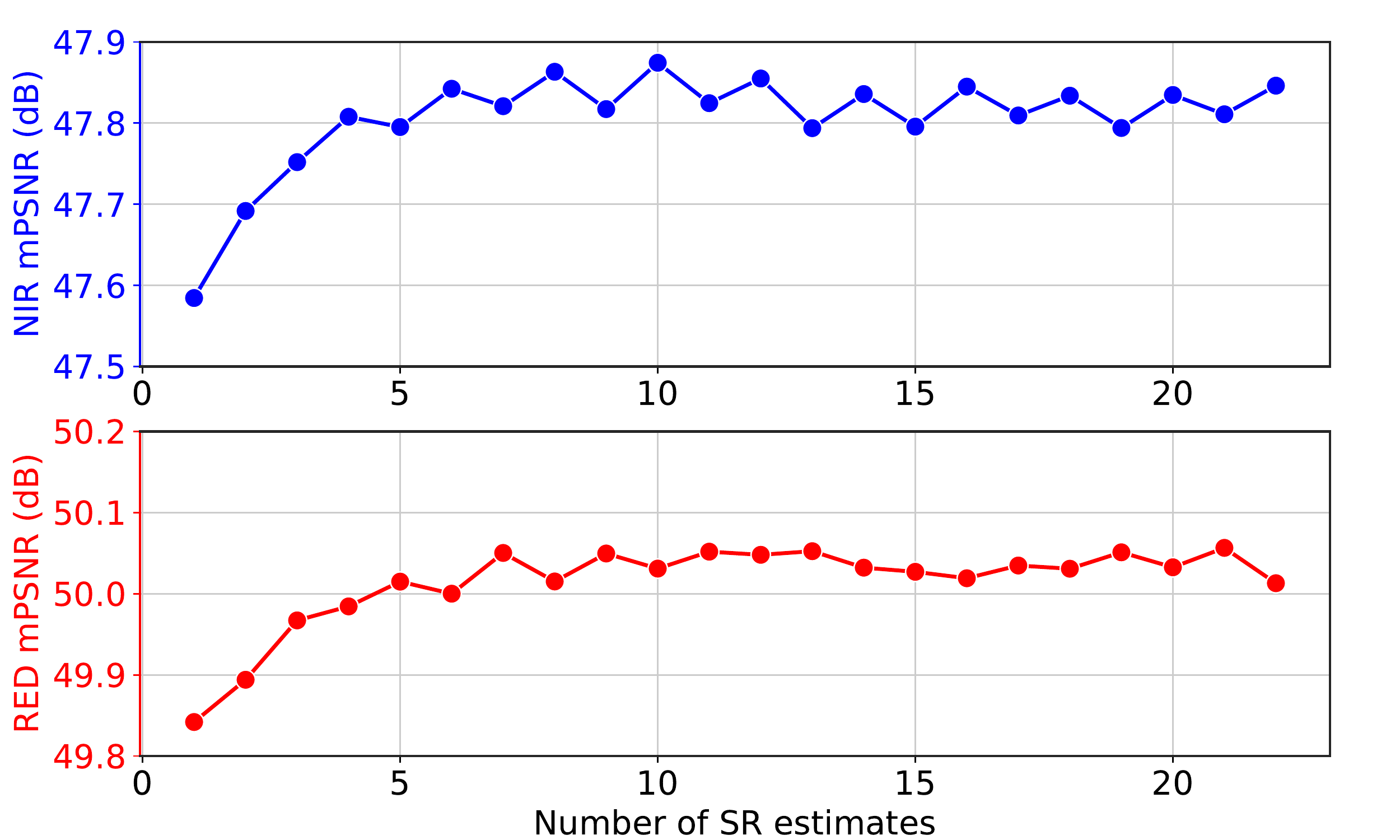}
    \caption{Effect of testing sliding window to deal with more than 9 LR images.}
    \label{fig:sliding}
\end{figure}

First, we want to assess the effectiveness of the sliding window procedure described in Sec.\ref{sec:testing} to account for more than 9 images for a given scene. Fig. \ref{fig:sliding} shows the mPSNR as function of the number of SR estimates used for computing the average. Notice that the mPSNR quickly saturates due to the lower quality of the images in the dataset (e.g., too many masked pixels). Nevertheless, averaging allow to achieve an mPSNR gain up to 0.3 dB over a single SR estimate on the NIR data and up to 0.2 dB on the RED  data. All the following results have been obtained with a sliding factor equal to 5. 

Then, we want to verify the effectiveness of the RegNet component of DeepSUM with respect to external registration of the images by means of cross correlation. This test should highlight the advantage of exploiting the feature space of the end-to-end trained network for the registration task. Hence, we compare two versions of our network: 
 
\begin{itemize}
    \item full network (SISRNet+RegNet+FusionNet);
    \item network without the RegNet block (SISRNet+FusionNet).
    We keep the registration filters but they are fixed to be a delta centered at the integer shift determined by maximum cross correlation on the ILR input images.
\end{itemize}
The full network outperforms the one without RegNet by 0.16 dB and 0.13 dB for the NIR and RED test sets, respectively, as shown in Table \ref{table:noregnet}. 
This is a significant margin and it is due to the fact that an inaccurate registration can be an important source of error for the SR reconstruction.

On the other hand, the full network, being trainable end-to-end, is able to exploit the feature space produced by SISRNet to provide a more accurate registration and help FusionNet to perform the feature merging task.
We remark that the full network and the reduced network have been trained independently.

\begin{table}[t]
\centering
\caption{Average mPSNR (dB) and SSIM - RegNet Performance}
\label{table:noregnet}
\begin{tabular}{lcc}
 & Proposed without RegNet & \textbf{Proposed with RegNet} \\ \hline
NIR & 47.68 / 0.98519 & \textbf{47.84 / 0.98578} \\ \hline
RED & 49.87 / 0.99038 & \textbf{50.00 / 0.99075} \\ \hline
\end{tabular}
\end{table}

\subsubsection{Comparison to State-of-the-Art}

\begin{table*}[t]
\centering
\caption{Average mPSNR (db) and SSIM}
\label{table:baselines}
\begin{tabular}{lcccccccc}
 & Bicubic & Bicubic+Mean & IBP \cite{IRANI1991231} & BTV \cite{1331445} & SISR & SISR+Mean & DUF \cite{Jo_2018_CVPR} & \textbf{DeepSUM} \\ \hline
NIR & 45.05/0.97654 & 45.69/0.97782 & 45.96/0.97960 & 45.93/0.97942 & 45.56/0.97938 & 46.41/0.98166 & 47.06/0.98417 & \textbf{47.84/0.98578} \\ \hline
RED & 47.61/0.98474 & 47.91/0.98507 & 48.21/0.98648 & 48.12/0.98606 & 48.20/0.98704 & 48.71/0.98787 & 49.36/0.98948 & \textbf{50.00/0.99075} \\ \hline
\end{tabular}
\end{table*}

We compare the proposed MISR technique to a number of alternatives based on deep learning and model-based methods:
\begin{enumerate}
    \item single image bicubic interpolation with least masked image (Bicubic);
    \item averaged bicubic interpolated and registered images (Bicubic+Mean);
    \item CNN-based SISR with least masked image;
    \item CNN-based SISR method shared across multiple images followed by registration and averaging (SISR+Mean);
    \item IBP \cite{IRANI1991231};
    \item BTV \cite{1331445};
    \item deep  learning  method  based  on  simultaneous motion compensation and interpolation developed for video (dynamic upsampling filters (DUF) network) \cite{Jo_2018_CVPR}.
\end{enumerate}
Table \ref{table:baselines} reports the results of the comparison. It can be noticed that the proposed method outperforms all the other methods.

For all these methods, we followed the same procedure for the data preparation: bicubic interpolation and registration by phase correlation algorithm, except for DUF that computes its own registration. For MISR methods we averaged the 5 SR estimates produced by the sliding window method to ensure a fair comparison with the proposed technique. 

Our IBP implementation takes as input an initial guess corresponding to our Bicubic+Mean baseline and the precomputed shifts related to the LR images using phase correlation algorithm. At each step, the LR images are estimated through the forward (HR to LR) imaging model and the error with respect to the actual LR images is back projected to the current SR image. We can observe that IBP improves over the Bicubic+Mean baseline but its performance is ultimately limited by its inability to deal with a complex and unknown degradation model.
BTV implementation takes the same initial guess and precomputed shifts as in IBP with the difference that at each iteration the cost function to minimize is a L1 norm plus the bilateral regularization term. 
BTV shows comparable performance with respect to IBP. BTV is slightly worse due to the L1 norm data fidelity that tends to be more robust to outliers but suboptimal with respect to the mPSNR metric.
The deep learning models show marked improvements over the Bicubic+Mean baseline. We consider two deep learning baselines (SISR only and SISR+Mean) that use the SISRNet architecture with the addition of a final layer projecting from the feature space to the image space, a residual connection from the (IRLR) bicubic image(s) and an increased number of parameters to roughly match the number of parameters of the full proposed architecture in order to ensure a fair comparison. The SISR+Mean result has been obtained by averaging 9 SISR images. Notice that SISR+Mean does not train the network by showing the averaged image to the loss function; it just uses the pretrained SISR network on multiple images and averages its outputs. The reason behind this choice is to provide a reference result to reader who might be interested in taking a state-of-the-art off-the-shelf SISR model, apply it to multiple images and then average the results. The comparison between SISR+Mean and the SISR only method is meant to highlight the large gain brought by exploiting both the spatial and temporal correlations, even if the LR images of a specific scene are taken under different conditions and might be wildly different from one another in terms of contrast, brightness and landscape due to temporal variations. Also, notice that SISR only is unable to improve over the simple Bicubic+Mean MISR on the NIR data. Instead, the comparison between DeepSUM and the SISR+Mean method shows the improvement brought by the introduction of FusionNet, which can exploit the slow fusion via 3D convolutions to find the best way to merge the image representations.

Another method chosen for comparison is the recent DUF network \cite{Jo_2018_CVPR}. This is one of the current state-of-the-art methods for video super-resolution. DUF network processes $N$ frames in order to compute local pixel-dependent dynamic filters that are later applied on the central frame to increase its resolution and compensate motion. The network has a residual branch estimating a residual image to increase sharpness of the final SR image.
The DUF network has been trained from scratch, maintaining the original structure and roughly the same number of learnable parameters with respect to our method for fair comparison. The only difference lies in using the loss function stated in Sec. \ref{subsec:loss} instead of the one used in the original paper (Huber loss).
Moreover, we always considered the first one among the 9 input LR images as central frame.
The performance is worse than our proposed method and we can deduce that it highly depends on the LR input image taken to apply the dynamic local filters. We cannot know in advance which is the LR image that is closer to the HR image due to change in brightness, landscape, weather, and clouds. Involving all the LR images for HR estimation is crucial to somehow average the differences across them and try to include as much information as possible in the final SR estimate.

For completeness, we report the score achieved by DeepSUM on the unreleased test set of the PROBA-V challenge. DeepSUM achieved a score equal to 0.9474466476281652, computed as the average ratio between the mPSNR of ESA's baseline and the mPSNR of the submitted images, over both RED and NIR data in the held-out test set.

\subsection{Qualitative results}
We present a set of qualitative comparisons on the RED and NIR images of our PROBA-V test set. 

\begin{figure*}[t]
  \centering
    \begin{minipage}[b]{\textwidth}
        \begin{minipage}[c]{0.15\textwidth}
        \includegraphics[width=\textwidth]{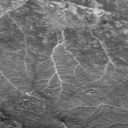}
        \end{minipage}
        \hfill
        \begin{minipage}[c]{0.15\textwidth}
        \includegraphics[width=\textwidth]{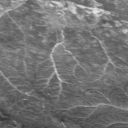}
        \end{minipage}
        \hfill
        \begin{minipage}[c]{0.15\textwidth}
        \includegraphics[width=\textwidth]{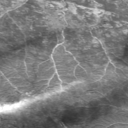}
        \end{minipage}
        \hfill
        \begin{minipage}[c]{0.15\textwidth}
        \includegraphics[width=\textwidth]{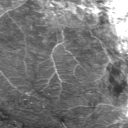}
        \end{minipage}
        \hfill
        \begin{minipage}[c]{0.15\textwidth}
        \includegraphics[width=\textwidth]{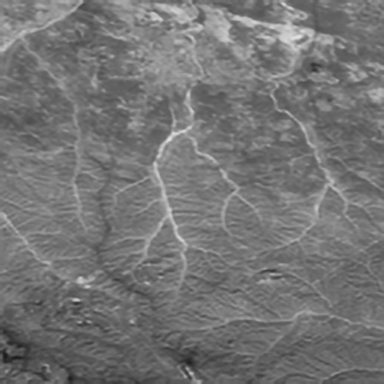}
        \end{minipage}
        \hfill
        \begin{minipage}[c]{0.15\textwidth}
        \includegraphics[width=\textwidth]{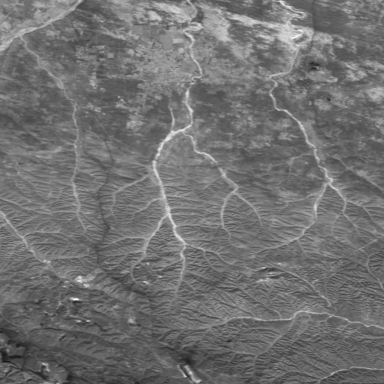}
        \end{minipage}
        \hfill
        \begin{minipage}[c]{0.041\textwidth}
        \includegraphics[width=\textwidth]{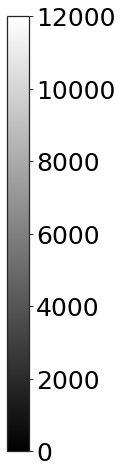}
        \end{minipage}
        
    \end{minipage}\\
    \caption{NIR band images (imgset0708). Left to right: 4 LR images, SR image reconstructed by DeepSUM and HR image.}
  \label{fig:many_lr_nir}
\end{figure*}

\begin{figure*}[t]
  \centering
    \begin{minipage}[b]{\textwidth}
        \begin{minipage}[c]{0.15\textwidth}
        \includegraphics[width=\textwidth]{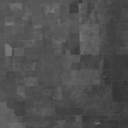}
        \end{minipage}
        \hfill
        \begin{minipage}[c]{0.15\textwidth}
        \includegraphics[width=\textwidth]{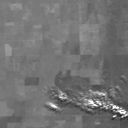}
        \end{minipage}
        \hfill
        \begin{minipage}[c]{0.15\textwidth}
        \includegraphics[width=\textwidth]{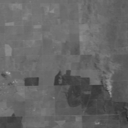}
        \end{minipage}
        \hfill
        \begin{minipage}[c]{0.15\textwidth}
        \includegraphics[width=\textwidth]{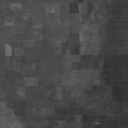}
        \end{minipage}
        \hfill
        \begin{minipage}[c]{0.15\textwidth}
        \includegraphics[width=\textwidth]{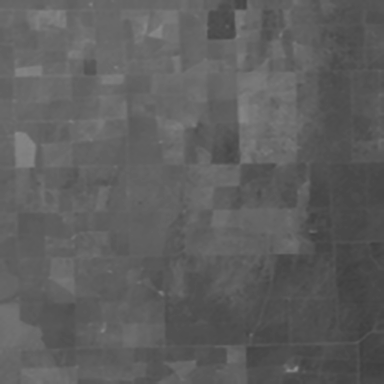}
        \end{minipage}
        \hfill
        \begin{minipage}[c]{0.15\textwidth}
        \includegraphics[width=\textwidth]{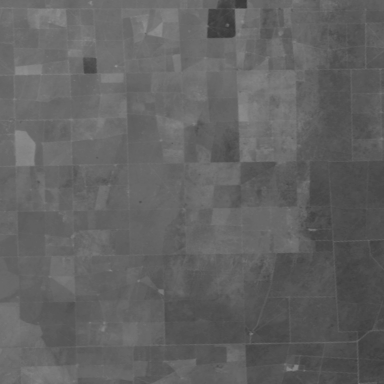}
        \end{minipage}
        \hfill
        \begin{minipage}[c]{0.041\textwidth}
        \includegraphics[width=\textwidth]{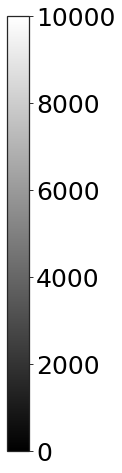}
        \end{minipage}
        
    \end{minipage}\\
    \caption{RED band images (imgset0103). Left to right: 4 LR images, SR image reconstructed by DeepSUM and HR image.}
  \label{fig:many_lr_red}
\end{figure*}

First of all, Figs. \ref{fig:many_lr_nir} and \ref{fig:many_lr_red} show the multitemporal variability among the LR images and between the LR set and the HR target for the NIR and RED bands, respectively.

\begin{figure*}[t]
  \centering
    \begin{minipage}[b]{\textwidth}
        \begin{minipage}[c]{0.23\textwidth}
        \includegraphics[width=\textwidth]{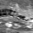}
        \end{minipage}
        \hfill
        \begin{minipage}[c]{0.23\textwidth}
        \includegraphics[width=\textwidth]{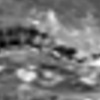}
        \end{minipage}
        \hfill
        \begin{minipage}[c]{0.23\textwidth}
        \includegraphics[width=\textwidth]{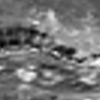}
        \end{minipage}
        \hfill
        \begin{minipage}[c]{0.23\textwidth}
        \includegraphics[width=\textwidth]{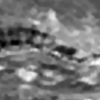}
        \end{minipage}
        \hfill
        \hspace{0.05\textwidth}
        
        \vspace{0.20cm}
        \begin{minipage}[c]{0.23\textwidth}
        \includegraphics[width=\textwidth]{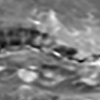}
        \end{minipage}
        \hfill
        \begin{minipage}[c]{0.23\textwidth}
        \includegraphics[width=\textwidth]{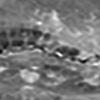}
        \end{minipage}
        \hfill
        \begin{minipage}[c]{0.23\textwidth}
        \includegraphics[width=\textwidth]{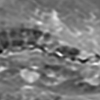}
        \end{minipage}
        \hfill
        \begin{minipage}[c]{0.23\textwidth}
        \includegraphics[width=\textwidth]{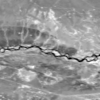}
        \end{minipage}
        \hfill
        \begin{minipage}[c]{0.05\textwidth}
        \includegraphics[width=\textwidth]{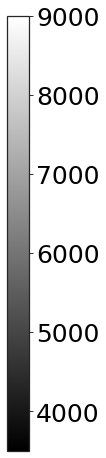}
        \end{minipage}
        
    \end{minipage}\\
    
    \caption{NIR band images (imgset0792). Top-Left to bottom-right: one among the LR images, Bicubic+Mean (47.71 dB / 0.98736), IBP (48.46 dB / 0.98919), BTV(48.12 dB / 0.98866), DUF (48.93 dB / 0.99028), proposed method without RegNet (50.71 dB / 0.99303), DeepSUM (50.82 dB / 0.99331), HR image}
  \label{fig:zoom_images_nir}
\end{figure*}

\begin{figure*}[t]
  \centering
    \begin{minipage}[b]{\textwidth}
        \begin{minipage}[c]{0.15\textwidth}
        \includegraphics[width=\textwidth]{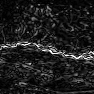}
        \end{minipage}
        \hfill
        \begin{minipage}[c]{0.15\textwidth}
        \includegraphics[width=\textwidth]{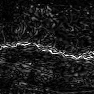}
        \end{minipage}
        \hfill
        \begin{minipage}[c]{0.15\textwidth}
        \includegraphics[width=\textwidth]{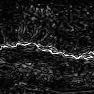}
        \end{minipage}\vspace{0.20cm}
        \hfill
        \begin{minipage}[c]{0.15\textwidth}
        \includegraphics[width=\textwidth]{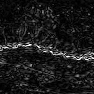}
        \end{minipage}
        \hfill
        \begin{minipage}[c]{0.15\textwidth}
        \includegraphics[width=\textwidth]{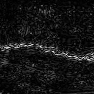}
        \end{minipage}
        \hfill
        \begin{minipage}[c]{0.15\textwidth}
        \includegraphics[width=\textwidth]{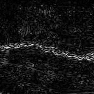}
        \end{minipage}
        \hfill
        \begin{minipage}[c]{0.035\textwidth}
        \includegraphics[width=\textwidth]{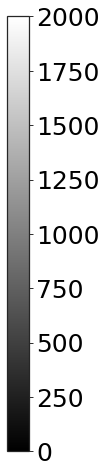}
        \end{minipage}
    \end{minipage}\\
    \caption{Absolute difference between SR image and HR image (NIR band). Left to right: Bicubic+Mean, IBP, BTV, DUF, proposed method without RegNet, DeepSUM}
  \label{fig:diff_nir}
\end{figure*}

\begin{figure*}[t]
  \centering
    \begin{minipage}[b]{\textwidth}
        \begin{minipage}[c]{0.23\textwidth}
        \includegraphics[width=\textwidth]{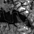}
        \end{minipage}
        \hfill
        \begin{minipage}[c]{0.23\textwidth}
        \includegraphics[width=\textwidth]{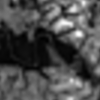}
        \end{minipage}
        \hfill
        \begin{minipage}[c]{0.23\textwidth}
        \includegraphics[width=\textwidth]{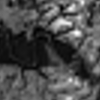}
        \end{minipage}
        \hfill
        \begin{minipage}[c]{0.23\textwidth}
        \includegraphics[width=\textwidth]{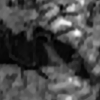}
        \end{minipage}
        \hfill
        \hspace{0.05\textwidth}
        
        \vspace{0.20cm}
        \begin{minipage}[c]{0.23\textwidth}
        \includegraphics[width=\textwidth]{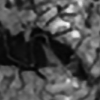}
        \end{minipage}
        \hfill
        \begin{minipage}[c]{0.23\textwidth}
        \includegraphics[width=\textwidth]{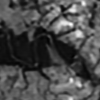}
        \end{minipage}
        \hfill
        \begin{minipage}[c]{0.23\textwidth}
        \includegraphics[width=\textwidth]{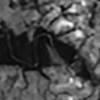}
        \end{minipage}
        \hfill
        \begin{minipage}[c]{0.23\textwidth}
        \includegraphics[width=\textwidth]{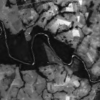}
        \end{minipage}
        \hfill
        \begin{minipage}[c]{0.05\textwidth}
        \includegraphics[width=\textwidth]{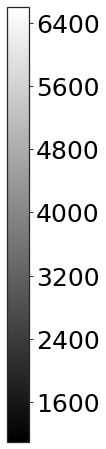}
        \end{minipage}
        
    \end{minipage}\\
    \caption{RED band images (imgset0184). Top-Left to bottom-right: one among the LR images, Bicubic+Mean (46.32 dB / ), IBP (46.52 dB / 0.97965), BTV (46.53 dB / 0.97983), DUF (47.64 dB / 0.98468), proposed method without RegNet (49.55 dB / 0.98886), DeepSUM (49.89 dB / 0.99041), HR image.}
  \label{fig:zoom_images_red}
\end{figure*}

\begin{figure*}[t]
  \centering
    \begin{minipage}[b]{\textwidth}       
        \begin{minipage}[c]{0.15\textwidth}
        \includegraphics[width=\textwidth]{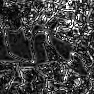}
        \end{minipage}
        \hfill
        \begin{minipage}[c]{0.15\textwidth}
        \includegraphics[width=\textwidth]{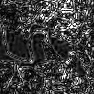}
        \end{minipage}
        \hfill
        \begin{minipage}[c]{0.15\textwidth}
        \includegraphics[width=\textwidth]{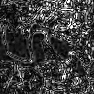}
        \end{minipage}\vspace{0.20cm}
        \hfill
        \begin{minipage}[c]{0.15\textwidth}
        \includegraphics[width=\textwidth]{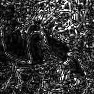}
        \end{minipage}
        \hfill
        \begin{minipage}[c]{0.15\textwidth}
        \includegraphics[width=\textwidth]{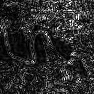}
        \end{minipage}
        \hfill
        \begin{minipage}[c]{0.15\textwidth}
        \includegraphics[width=\textwidth]{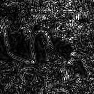}
        \end{minipage}
        \hfill
        \begin{minipage}[c]{0.035\textwidth}
        \includegraphics[width=\textwidth]{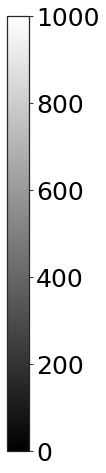}
        \end{minipage}
        
    \end{minipage}\\
    \caption{Absolute difference between SR image and HR image (RED band). Left to right: Bicubic+Mean, IBP, BTV, DUF, proposed method without RegNet, DeepSUM.}
  \label{fig:diff_red}
\end{figure*}

Figs. \ref{fig:zoom_images_nir} and \ref{fig:zoom_images_red} show a visual comparison between the SR images reconstructed by the various methods for the NIR and RED bands, respectively. It can be noticed that our proposed method produces visually more detailed images, recovering finer texture and sharper edges. In order to help visualization, Figs. \ref{fig:diff_nir} and \ref{fig:diff_red} report the absolute difference between the HR target and the SR reconstructions for the various methods after registration and compensation for absolute brightness variations (as in the mPSNR computation).

\section{Conclusion}
\label{sec:conclusions}
In this paper we have introduced DeepSUM, one of the first CNN architectures to deal with super-resolution from multitemporal remote sensing images. We showed that the proposed deep learning framework can successfully deal with complex degradation and temporal variation models and provide state-of-the-art performance, resulting as the best method in the PROBA-V SR challenge.
Future work may focus on integrating non-local features in the network, e.g., by using graph-convolutional architectures \cite{ValsesiaICIP19}, a kind of convolution that draws from ideas in graph signal processing \cite{shuman2013emerging,Valsesia2019Sampling}.

\ifCLASSOPTIONcaptionsoff
  \newpage
\fi

\begin{IEEEbiography}[{\includegraphics[width=1in,height=1.25in,clip,keepaspectratio]{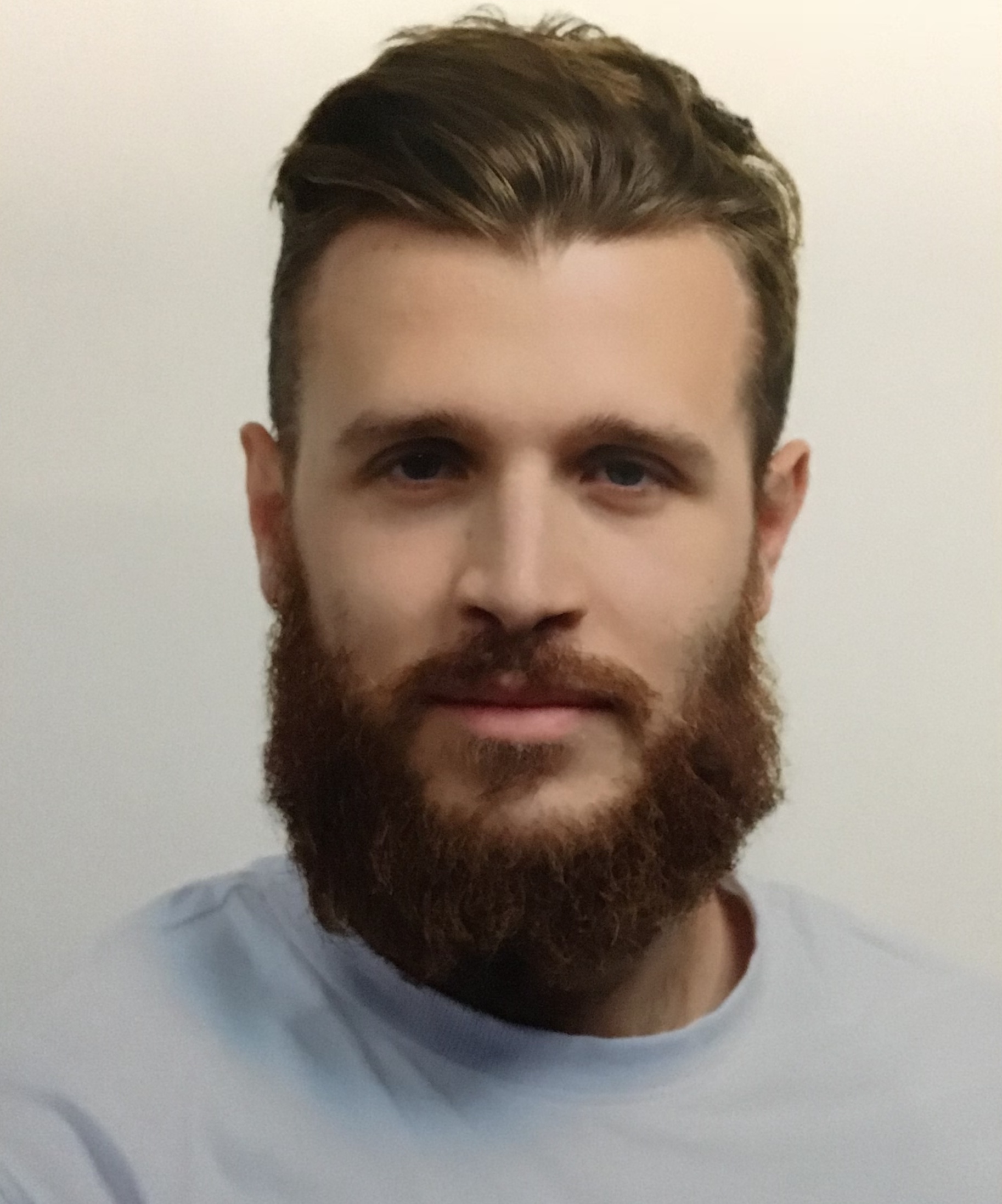}}]{Andrea Bordone Molini} received the M.Sc. in computer networks engineering from Politecnico di Torino, Turin, Italy, and a M.Sc. in Communications Systems Security from T\'el\'ecom ParisTech, Paris, France as part of a double degree program in 2016. He joined the SmartData@Polito research center where he is pursuing the Ph.D. degree. His current research interests include deep learning applied to image processing in particular in the fields of super-resolution and denoising. \end{IEEEbiography}

\begin{IEEEbiography}[{\includegraphics[width=1in,height=1.25in,clip,keepaspectratio]{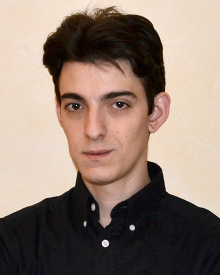}}]{Diego Valsesia} (S'13-M'17) received the Ph.D. degree in Electronic and Communication Engineering from the Politecnico di Torino, Turin, Italy, in 2016 and the M.Sc. degrees in Telecommunications Engineering from Politecnico di Torino and Electrical and Computer Engineering from the University of Illinois at Chicago, Chicago, IL, in 2012 and 2013 respectively. He is currently an Assistant Professor at the Department of Electronics and Telecommunications (DET), Politecnico di Torino. His main research interests include compression of remote sensing images, compressed sensing, and deep learning.\end{IEEEbiography}

\begin{IEEEbiography}[{\includegraphics[width=1in,height=1.25in,clip,keepaspectratio]{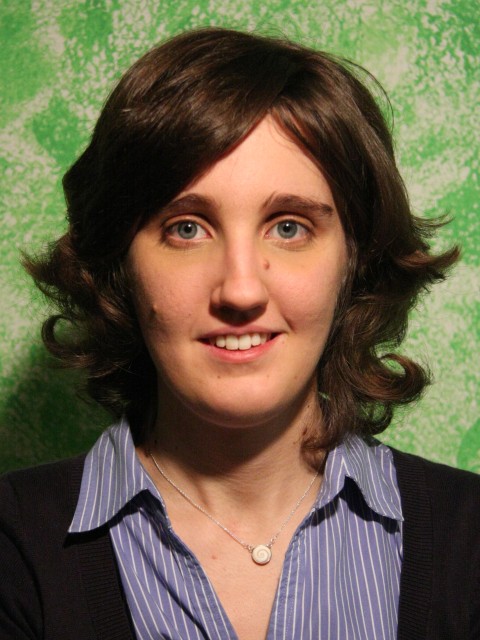}}]{Giulia Fracastoro} (S'13-M'17) Giulia Fracastoro received the Ph.D. degree in Electronic and Telecommunications Engineering from Politecnico di Torino, Turin, Italy, in 2017. During 2016, she was a visiting student at the Signal Processing Laboratory at EPFL, working on graph learning for image compression. She is currently an Assistant Professor with the Department of Electronics and Telecommunications (DET), Politecnico di Torino. Her research interests include graph signal processing, image processing, and deep learning.\end{IEEEbiography}

\begin{IEEEbiography}[{\includegraphics[width=1in,height=1.25in,clip,keepaspectratio]{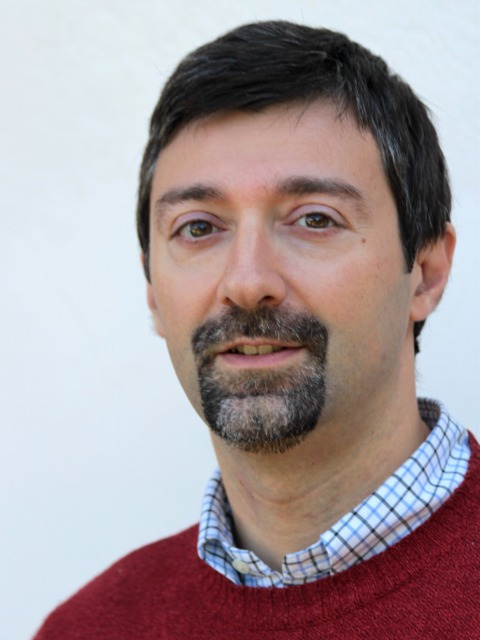}}]{Enrico Magli} (S'97-M'01-SM'07-F'17) received the M.Sc. and Ph.D. degrees from the Politecnico di Torino, Torino, Italy, in 1997 and 2001, respectively. He is currently a Full Professor with Politecnico di Torino, Torino, Italy. His research interests include compressive sensing, image and video coding, and vision. He is an Associate Editor of the IEEE TRANSACTIONS ON CIRCUITS AND SYSTEMS FOR VIDEO TECHNOLOGY and the EURASIP Journal on Image and Video Processing, and a former Associate Editor of the IEEE TRANSACTIONS ON MULTIMEDIA. He is a Fellow of the IEEE, and has been an IEEE Distinguished Lecturer from 2015 to 2016. He was the recipient of the IEEE Geoscience and Remote Sensing Society 2011 Transactions Prize Paper Award, the IEEE ICIP 2015 Best Student Paper Award (as senior author), and the 2010 and 2014 Best Associate Editor Award of the IEEE TRANSACTIONS ON CIRCUITS AND SYSTEMS FOR VIDEO TECHNOLOGY.
\end{IEEEbiography}

\end{document}